\documentclass[11pt,twoside]{article}
\usepackage{asp2010}

\resetcounters

\begin{document}

\title{Bring out your codes! Bring out your codes! (Increasing Software Visibility and Re-use)}
\author{Alice Allen$^1$, Bruce Berriman$^{2,3}$, 
Robert Brunner$^4$, Dan Burger$^5$, Kimberly DuPrie$^1$, Robert J. Hanisch$^{6,3}$, Robert Mann$^7$, Jessica Mink$^8$, Christer Sandin$^9$, Keith Shortridge$^{10}$, and Peter Teuben$^{11}$
\affil{$^1$Astrophysics Source Code Library}
\affil{$^2$Infrared Processing and Analysis Center, California Institute of Technology}
\affil{$^3$Virtual Astronomical Observatory}
\affil{$^4$University of Illinois}
\affil{$^5$Vanderbilt University}
\affil{$^6$Space Telescope Science Institute}
\affil{$^7$University of Edinburgh/Editor, Astronomy \& Computing}
\affil{$^8$Harvard-Smithsonian Center for Astrophysics}
\affil{$^9$Leibniz Institute for Astrophysics Potsdam}
\affil{$^{10}$Australian Astronomical Observatory}
\affil{$^{11}$Astronomy Department, University of Maryland}}

\begin{abstract}

Progress is being made in code discoverability and preservation, but as discussed at ADASS XXI, many codes still remain hidden from public view. With the Astrophysics Source Code Library (ASCL)\footnote{http://ascl.net/} now indexed by the SAO/NASA Astrophysics Data System (ADS), the introduction of a new journal, {\em Astronomy \& Computing}, focused on astrophysics software, and the increasing success of education efforts such as Software Carpentry and SciCoder, the community has the opportunity to set a higher standard for its science by encouraging the release of software for examination and possible reuse. We assembled representatives of the community to present issues inhibiting code release and sought suggestions for tackling these factors.

The session began with brief statements by panelists; the floor was then opened for discussion and ideas. Comments covered a diverse range of related topics and points of view, with apparent support for the propositions that algorithms should be readily available, code used to produce published scientific results should be made available, and there should be discovery mechanisms to allow these to be found easily. With increased use of resources such as GitHub (for code availability), ASCL (for code discovery), and a stated strong preference from the new journal {\em Astronomy \& Computing} for code release, we expect to see additional progress over the next few years.
\end{abstract}

\section{Introduction}

This Birds of a Feather (BoF) session was held to gather ideas on how to better make astronomical software discoverable and preserve it to improve the transparency of research, ensure reproducibility, and advance numerical methods in the field. How to handle software is often discussed at ADASS, and this BoF builds directly on Teuben et al. (2012). The authors feel that recent changes in the field create an opportunity to put the community on a path for permanent change regarding software discoverability, sharing, and advancement of these methods.

\section{Efforts in the Community}

The astronomy community has begun to recognize the need to and benefits of publishing and sharing software. A number of efforts are building greater cooperation and actively working to improve the way astrophysics is done, and have started to have an impact on the development, visibility, and preservation of codes. These efforts include:

\begin{itemize}
\item Blogs such as Astronomy Computing Today\footnote{http://astrocompute.wordpress.com/} and AstroBetter,\footnote{http://www.astrobetter.com/} devoted in part or wholly to software topics and doing things better.
\item Projects such as Software Carpentry\footnote{http://software-carpentry.org/}  and SciCoder\footnote{http://www.scicoder.org/} to improve coding skills.
\item Increasing efforts to recognize the role of the astronomical software professional in advancing the field through the development of astroinformatics conferences, coursework, and code citation.
\item Expansion of the ASCL, indexing of its entries by ADS, and ADS's exploration of linking papers to code entries and code entries to papers.
\item Collaborative coding efforts such as AstroPy.\footnote{http://www.astropy.org/}
\item Social software bringing astronomers together in previously unprecedented ways ({\em i.e.}, Astronomers' Facebook group, AstroShare Google group\footnote{https://groups.google.com/forum/?fromgroups\#!forum/AstroShareGroup}).
\item Launch of a journal, {\em Astronomy \& Computing}, devoted to the development and use of software methods in astronomy.
\end{itemize}

\section{Barriers to Code Discoverability}

Despite progress, significant barriers still exist which inhibit code release; many of these have been discussed in the literature and at ADASS. Reasons developers do not release codes include:

\begin{itemize}
\item Code ``messiness" (Barnes, 2010).
\item University policies that prohibit distribution of intellectual property.
\item Lack of documentation and examples.
\item Perceived lack of suitability for sharing, as the code may have a narrow focus and/or seem too trivial to share.
\item Protection of proprietary processes useful for future funding of author.
\item Code release not firmly established as a standard practice.
\item Lack of incentive/little or no perceived upside to releasing code.
\end{itemize}

\section{Discussion Questions}

The focus of the BoF was to discuss and solicit answers to the following questions:

\begin{itemize}
\item How do we ensure code release is recognized as an essential part of assuring reproducibility of research?
\item How can the community change the culture so coders will release their programs?
\item What can we do to ensure developers receive credit for writing and releasing their software, and encourage them to release it even if it's ``messy'' code?
\item How do we reduce expectations of support when a developer does not wish to (or cannot) take on that role after program release?
\item What role might journal publishers and funding agencies have in furthering code release, and how can the community influence them to take on that role?
\item How can universities be convinced to change policies that prohibit software publication?
\item Can funding agencies and publishers encourage documentation of programs, and if so, how?
\end{itemize}

\section{Discussion}

Discussion was spirited and wide-ranging, with several themes emerging from it: 
\begin{itemize}
\item There is no ``one size fits all" solution. There are many different classes of code being written, from one-off tests to the codes used to produce published results.
\item It is important to distinguish between code and algorithms; often people care more about the algorithms, and while code may embody an algorithm it can also obscure it.
\item  Insisting that every last piece of code be made available sounds unrealistic, but code used to produce published results really matters. Some journals and funding bodies are starting to see such code as an integral part of the work, so as something that needs to be available.
\item Versions of code become important for understanding its evolution and in order to know just which version produced a published result.
\item It is important to distinguish between accessibility and discoverability, although both matter. Sites such as ASCL address the discoverability problem.
\item Some people, particularly with a computer science background, are comfortable with putting code into GitHub from the very start which solves both the accessibility and the versioning problem.
\end{itemize}

Releasing software increases transparency, though some are reluctant to release code in part because of a perceived need to be available to support the code. This could be mitigated with the implementation of the Community Research and Academic Programming License, or CRAPL\footnote{http://matt.might.net/articles/crapl/}. Looking at messy codes can be instructive; software authors could partner with computer science students to clean code up. Science can be advanced by improving the skills of code authors in areas of testing (creating and publishing test cases), clean coding, version control, and documentation. Though these did not come up in the BoF, suggestions for improvement can be found in Aruliah (2012). 

In the longer term, astronomy will advance faster with greater transparency of software and adoption of open source licenses, as others can improve and extend work that has already been done; methods already exist for publishing codes and managing code changes/versions and releases. The science can be improved by improving software distribution, which can be aided by the use of social coding sites.

\section{Conclusions}

Cultural change will come about fastest when agencies demand code release and as publications encourage or insist on it. Cost is a factor -- ideally, a grant would cover the cost of a professional software engineer to write any necessary code, but we all know that isn't always possible. Therefore, science is better served by scientists releasing their codes and ideally learning better coding and code release practices.

\end{document}